# LITHIUM ABUNDANCE AND SURFACE MAGNETIC FIELDS: NEW CONSTRAINTS ON MAGNETIC MODELS OF M DWARFS

JAMES MACDONALD AND D. J. MULLAN

Department of Physics and Astronomy, University of Delaware, Newark, DE 19716; jimmacd@udel.edu, mullan@udel.edu

## ABSTRACT

Precision modeling of M dwarfs has become worthwhile in recent years due to the increasingly precise values of masses and radii which can be obtained from eclipsing binary studies. Torres (2013) has identified 4 prime M dwarf pairs with the most precise empirical determinations of masses and radii. The measured radii are consistently larger than standard stellar models predict. We have previously modeled M dwarfs in the context of a criterion due to Gough & Tayler in which magnetic fields inhibit the onset of convection according to a physics-based prescription. New constraints on the models of M dwarfs are now provided by measurements of lithium abundances. The key aspect of Li in terms of setting constraints on magnetic modeling is that Li burning starts at $T = 2.5$ MK, and temperatures of just such magnitude are associated with the base of the convection zone: magnetic inhibition of convective onset can shift this base slightly closer to the surface, i.e. to slightly lower temperatures, thereby reducing the amount of Li depletion compared to a non-magnetic model. In the present paper, we consider how our magneto-convection models handle the new test of stellar structure provided by Li measurements. Among the prime systems listed by Torres, we find that plausible magnetic models work well for CM Dra and YY Gem but not for CU Cnc. (The fourth system in Torres's list does not yet have enough information to warrant magnetic modeling.) For CU Cnc, we suggest that the observed lithium may have been accreted from a circumstellar disk. We find that our magneto-convection models of CM Dra, YY Gem and CU Cnc yield results which are consistent with the observed correlation between magnetic flux and X-ray luminosity.

## 1. INTRODUCTION

In a recent review, Torres (2013) has singled out 4 double line eclipsing M dwarf binaries as systems in which mass and radius have been measured with the highest precision: CM Dra, YY Gen, CU Cnc and GU Boo. In each case the measured radii are 5 – 10% larger than predicted by standard stellar evolution models. The excesses in radii are statistically significant, possibly by as much as $20\sigma$ in the case of YY Gem (Torres & Ribas 2002). These M dwarfs currently provide the most stringent challenges to the theoretical goal of precision stellar modeling. All four of these binary systems have short orbital periods, from 0.48 to 2.77 days. As a result of tidal forces in these short period binaries, the stellar spins are likely to be synchronized with the orbital motion, leading to rapid rotation and associated magnetic dynamo action.

*1.1. Magnetic effects in stellar models*

Two reasons have been proposed to explain why magnetic fields can lead to an increase in the radius of a low mass star compared to a non-magnetic star with otherwise identical properties.

First, magnetic fields can alter the global internal structure of a star because the presence of a vertical magnetic field in a star inhibits the onset of convective energy transport (Gough & Taylor 1966: GT). It is important to emphasize that there is nothing *ad hoc* about the criterion derived by GT for magnetic inhibition of convection: the criterion was derived by applying a widely used energy principle (Bernstein et al. 1958) to an electrically conducting fluid. The principle is based on perturbing an equilibrium configuration by moving an element of fluid through a vector distance $\xi$ from its initial position, and then evaluating the formal change in potential energy $\delta W$. The system is unstable if $\delta W$ turns out to be negative, and stable if $\delta W$ is positive for all perturbations which satisfy certain boundary conditions. When the GT criterion was incorporated into a stellar structure code for low mass main sequence stars (Mullan & MacDonald 2001: MM01), the results indicated that the radius of a magnetic star exceeds the value predicted for a non-magnetic star using the same code.

Second, magnetic fields can give rise to dark spots on the surface of a star, thereby hindering radiative losses from the surface: this process also results in increased radii (Spruit 1992; Chabrier, Gallardo, & Baraffe 2007).

When comparing a measured stellar radius with the radius predicted by a model of the correct mass, the conclusion regarding an excess in the radius can be considered reliable only in cases where the age and composition of the star in question are well constrained. Incorrect input data to a model can lead to unreliable conclusions. As an illustration, Torres (2013) has shown that, if [Fe/H] is *assumed* to be equal to +0.5, then the observed radii of the components of CM Dra *could* be replicated by the standard evolution models of Dotter et al (2008) without the need to invoke magnetic effects of any kind. However, CM Dra is known to be a Population II object, and recent determinations of the heavy element abundances suggest metal abundances which are 6-7 times smaller than the value of [Fe/H] needed by non-magnetic models: reports from Kuznetsov et al. (2012) and Terrien et al. (2012) indicate that [Fe/H] = -0.3. Using this smaller metal abundance, standard (non-magnetic) stellar evolution models predict radii which are definitely too small to match the CM Dra observations within the error bars. In a study of CM Dra which was undertaken before the metal abundance [Fe/H] = -0.3 was published, MacDonald & Mullan (2012: hereafter MM12) obtained a model using an intermediate metal abundance [Fe/H] = 0.04. With an age of 4 Gyr, MM12 showed that the radius and luminosity measurements of CM Dra A can be reproduced by stellar models in which a magnetic field inhibits convective energy transport and causes a dark spot covering 17% of the surface area. In these models, the vertical magnetic field strength at the surface of CM Dra A was found to be 500 G. Also in these models, based on estimates of equipartition between magnetic energy density and the kinetic energy density of rotational motion (leading to limits on field strengths which might be generated by a dynamo), the field strength inside the MM12 model of CM Dra A was not allowed to exceed a "ceiling" of $B_{ceiling}$ = 1 MG.

In the present paper, we first re-visit our GT modeling of CM Dra in the light of the recent abundance measurements. We then apply our GT models to two more of the most challenging M dwarf binaries highlighted by Torres (2013): YY Gem and CU Cnc. For both of these systems, information about age and [Fe/H] is available, and so a concerted modeling effort aimed at these two systems is now worth undertaking. The fourth candidate in Torres' list, GU Boo, remains ambiguous as regards age and/or composition (Lopez-Morales & Ribas 2005): it does not yet appear to be suitable for precision modeling.

*1.2. A new constraint on precision modeling of M dwarfs*

There is one further observational characteristic of both YY Gem and CU Cnc which distinguishes them in an important way from stars to which the GT criterion has previously been applied: the element lithium has been reported in the spectra of both systems. This imposes a new constraint on models, over and above the constraints of radius, mass, and luminosity. Specifically, the presence of Li requires that the temperature at the bottom of the outer convection zone has not exceeded 2.5 MK (Bodenheimer 1965) in the course of stellar evolution. In a star such as the Sun, where the convection zone has a maximum temperature of about 2 MK, lithium has not been depleted to a great extent. But for stars of sub-solar mass, the convection zone penetrates deeper than in the Sun. As a result, complete depletion of lithium becomes in principle possible. This is especially true if the star has low enough mass that convection can penetrate all the way to the center of the star: the latter circumstance probably applies to both components of CM Dra. The systems YY Gem and CU Cnc of prime interest to us in this paper have the important characteristic that all 4 components are intermediate in mass between the Sun and the mass of a completely convective star. The latter is typically less than about 0.35 $M_\odot$ while the stars in YY Gem both have masses around 0.6 $M_\odot$ and the stars in CU Cnc both have masses around 0.4 $M_\odot$. In view of this, the components of YY Gem and CU Cnc are prime candidates for exploring how sensitive the Li depletion is to relatively small changes in the temperature at the base of the convection zone. In this regard, models in which the physics of convective onset is controlled by magnetic fields (e.g. the GT criterion) would appear to be especially suited to modeling the temperatures at the relevant depths, i.e. at the base of a stellar convection zone.

In the present paper, we apply the GT criterion to YY Gem and CU Cnc. We shall find that the Li information constrains the value of the magnetic field "ceiling" in the stellar interior. A major new result of the present paper is that, in the context of GT modeling, the Li constraint leads to strikingly different results in our YY Gem models from those in our CU Cnc models. In one case (YY Gem), our GT models succeed in fitting both radius and Li abundance only if the field $B_{ceiling}$ does *not* exceed a certain *upper limit*. In particular, we can obtain fits to YY Gem for $B_{ceiling}$ as low as 1 MG, just as we obtained for CM Dra. But in contrast, when we attempt to model CU Cnc and fit the Li abundance, our GT models require that magnetic fields must inhibit convection very deep into the star: to achieve such inhibition, the models require that $B_{ceiling}$ must *exceed* a certain *lower limit*. And the lower limit is very large: 100 MG, much larger than the field which could reasonably expected to be generated inside a star based on equipartition arguments. It is difficult to consider fields of >100 MG as being plausible in M dwarfs. This distinction between YY Gem and CU Cnc in terms of interior field strengths is even more striking when we note that YY Gem has an orbital period which is 3-4 times faster than that of CU Cnc: this suggests that, other things being equal, the fields in YY Gem might be expected to be *stronger* than those in CU Cnc. But our models suggest precisely the opposite.

In an attempt to resolve this difficulty with interior models, we consider a scenario in which the presence of Li in CU Cnc might have nothing to do with the *internal* stellar structure. Instead, we ask: might the observed Li be due to accretion from a circumstellar disk? We argue that such a scenario is not inconsistent with infrared data. Once the Li constraint on the interior model is removed, we find that models with $B_{ceiling}$ = 1 MG can also be found which fit CU Cnc.

Each of our GT models provides naturally a value of the vertical field strength on the surface of each star. We show that for CM Dra, YY Gem, and CU Cnc, these surface fields are consistent with the empirical relation which has been reported between X-ray luminosity and measured surface magnetic flux (Fisher et al. 1998; Pevtsov et al., 2003; Feiden & Chaboyer 2013).

The plan of the paper is as follows. In Section 2, we summarize our modeling approach. In Section 3, we re-visit the magnetic modeling of CM Dra, updating our earlier models to incorporate the more recent metal abundances. In Sections 4 and 5, we obtain magneto-convection models for YY Gem and CU Cnc respectively, including the lithium data. In Section 6 we include a discussion and summary: there, we suggest that, in view of our lack of success in obtaining a magneto-convection model of CU Cnc which also replicates the Li abundance with plausible field strengths, it is worthwhile to explore possibilities for alternative explanations for the Li data.

2. MODELING TECHNIQUE

*2.1. Input physics*

Our stellar structure and evolution code is described in MacDonald & Mullan (2013) and references therein. For the calculations described here we use the OPAL equation of state (Rogers & Nayfonov 2002). The Eddington approximation is used to set the outer boundary conditions. Specifically, we determine the temperature and pressure at optical depth 0.5. This choice of optical depth gives radii of non-magnetic lower main sequence models close to those of models that use the NextGen atmospheres (Hauschildt et al. 1999; Allard et al. 2000) to determine the outer boundary condition. We use the Eddington boundary conditions for our magnetic models because i) the atmosphere models do not include magnetic effects, and ii) they are available only for a single mixing length ratio.

*2.2. Magneto-convection*

Our model for magneto-convection has been developed in a sequence of papers beginning with MM01. It is based on a criterion derived by Gough & Tayler (1966) who found that convective stability in the presence of a uniform vertical magnetic field $B_v$ frozen in an ideal gas of pressure $P_{gas}$ and infinite electrical conductivity is ensured as long as $\nabla_{rad}$ does not exceed $\nabla_{ad} + \delta$, where the magnetic inhibition parameter $\delta$ is defined (in Gaussian cgs units) by

$$\delta = \frac{B_v^2}{B_v^2 + 4\pi\gamma P_{gas}}, \quad (1.1)$$

where $\gamma$ is the first adiabatic exponent.

The Gough – Tayler criterion may not directly apply to convection in a cool magnetic dwarf for two reasons: 1) The gas is far from ideal and 2) the electrical resistivity can be much higher than for fully ionized plasma. The finite magnetic diffusivity allows fluid to move across magnetic field lines, and this can weaken the ability of the magnetic field to hinder thermal convection. To take into account non-ideal thermodynamic behavior and the effects of finite electrical conductivity, Mullan & MacDonald (2010) modified the Gough & Tayler criterion for convection to

$$\nabla_{rad} - \nabla_{ad} > \Delta \equiv \frac{\delta}{\theta_e} \min\left(1, \frac{2\pi^2 \gamma \kappa}{\eta \alpha^2}\right). \quad (1.2)$$

Here the non-ideal behavior is accounted for by the inclusion of a dependence on the thermal expansion coefficient $\theta_e = -\partial \ln \rho / \partial \ln T |_P$. The factor $\min(1, 2\pi^2 \gamma \kappa / \eta \alpha^2)$, in which $\kappa$ is the thermal conductivity and $\eta$ is the magnetic diffusivity, gives a transition to a generalization (Cowling 1957) of the criterion derived by Chandrasekhar (1961, and references therein) for the onset of thermal convection in the presence of a vertical magnetic field in a thermally conducting and magnetically diffusive incompressible fluid. Our method of calculation of the electrical conductivity is given in MacDonald & Mullan (2009). We shall refer to eq. (1.2) as the GTC criterion when the finite conductivity correction factor is included and as the GT criterion when the finite conductivity correction is omitted.

Although the GTC criterion gives the correct qualitative behavior in the limits of $\kappa/\eta$ being large or small, we do not claim that it is precise. Uncertainties lie in the choice of the numerical factor $2\pi^2/\alpha^2$, and an additional multiplicative factor, $f_{ec}$, could be included to allow for these uncertainties in the effects of electrical conductivity. However, it is important to keep in mind the following two observational properties of the stars which are being discussed here. First, cool star-spots are present in many M dwarfs, including all 4 of the objects high-lighted by Torres (2013): Kron (1952) reported the discovery of spots in the components of YY Gem, while Lacy (1977), Ribas (2003), and Lopez-Morales & Ribas (2005) did the same for CM Dra, CU Cnc, and GU Boo respectively. Second, many M dwarfs exhibit significant chromospheric activity, which, in the presence of strong enough heating, can produce emission components in the Balmer lines (Cram & Mullan 1979). Cool dwarf stars which exhibit Balmer emission lines are denoted by spectral type dMe or MVe. Thus, Gershberg (2002), in his catalog of flare stars, lists YY Gem as dM1e+dM1e (p. 658), CM Dra as dM4e/dM4.5 (p. 662), CU Cnc as M5e/M3.5, indicating strong chromospheric heating. (GU Boo is not yet known to be a flare star, and therefore does not appear in Gershberg's list.) The existence of spots and chromospheres indicates clearly that there must be significant coupling between matter and magnetic field in active M dwarfs stars: the presence of cool spots requires fields to dominate over matter (impeding convective flows: see Mullan 1974), while the presence of chromospheres requires matter to dominate over fields (stressing the fields and leading to enhanced local energy dissipation: see Mullan 2010). The presence of both spot and chromospheric phenomena in the 4 stars of interest to us here suggests that matter and fields are tied closely together, analogous to the limit of ideal conductivity. As a result, in the 4 stars mentioned by Torres (2013), finite conductivity effects may be negligible. Hence in this work, we have used the GT criterion, i.e. eq. (1.1) above, without the finite conductivity correction factor, as the criterion for onset of convective stability.

To determine the convective energy flux, we replace $\nabla_{ad}$ by $\nabla_{ad} + \Delta$ everywhere it appears in the mixing length theory. Our specific choice of mixing length theory is that of Mihalas (1978), which is the same as that of Böhm-Vitense (1958) but modified to include a correction to radiative losses from convective elements when they are optically thin.

The magnetic inhibition parameter $\delta$ is a local variable: in general, its numerical value may vary as a function of radial position in a star, and the question arises as to the appropriate choice for the radial profile of $\delta(r)$. Here we make use of dynamo concepts, as reported in our modeling of CM Dra (MM12), to set the profile. We take $\delta$ to be constant from the surface to the radial location at which the local magnetic field strength reaches a prescribed value (the 'ceiling'). At radial locations which are deeper than that depth, we hold the field strength constant at $B_{ceiling}$: as a result of this restriction, the numerical value of $\delta$ ($\sim B^2/p_{gas}$) decreases to very small values in the inner regions of the stellar model, where $p_{gas}$ continually increases towards the center of the star. Empirical support for our imposition of reduced values of $\delta$ in the deeper layers of one particular star (the Sun) has been provided recently by analysis of alterations in p-mode frequencies between solar minimum and solar maximum (Mullan, MacDonald & Rabello-Soares 2012).

*2.3. Star spots*

Our treatment of star spots is the same as in MM12. The influence of star spots on internal stellar structure has been reviewed by Spruit (1992). The blocking effects of spots is modeled by modifying the surface boundary condition to

$$L = 4\pi R^2 (1 - f_s) \sigma T_u^4, \quad (1.3)$$

where $f_s$ is the effective fraction of the surface covered by spots, which are assumed to be completely dark, and $T_u$ is the surface temperature of the immaculate (unspotted) surface, which is set equal to the model temperature at optical depth 2/3. For fully convective stars, spots result in a reduction in luminosity given by $\Delta L/L \approx -f_s$, with much smaller relative changes in $R$ and $T_u$. Note that because $T_u$ does not change significantly in the presence of a spot, any significant spot coverage will reduce $T_{eff}$ for fully convective stars according to the expression $T_{eff}^4 = (1 - f_s) T_u^4$.

3. CM DRACONIS REVISITED

CM Draconis (CM Dra) is an eclipsing binary, of orbital period $P_{orb}$ = 1.27 d, containing two dwarfs with mid-M spectral types. Spectroscopic data which span appropriate regions of the spectrum indicate that the Balmer lines are in emission, thereby meriting a dMe classification for one or both components. Thus, spectral types are listed as dM3-4e (Eggen & Sandage 1967), dM5e (Greenstein 1969), and dM4e (Lacy 1977). In what might appear, at first sight, to be a contradiction to the dMe classification, Morales et al. (2009) refer to the system as "two dM4.5 stars", without adding the "e". However there is no contradiction: the spectral range used by Morales et al. (spanning 45 Å near $\lambda$5187 Å) does not include any Balmer lines. Therefore, on the basis of those data, nothing can be said one way or another about Balmer line emission in CM Dra. But in fact, Morales et al. state that Barnard's star "with a spectral class M4Ve provides a close match to the spectral type of CM Dra". Thus, the dMe classification, along with its indication of strong chromospheric activity (and therefore close field-material coupling), is appropriate for CM Dra. The masses and radii of the binary components are $M_A$ = 0.23102 ± 0.00089 $M_\odot$, $R_A$ = 0.2534 ± 0.0019 $R_\odot$ and $M_B$ = 0.21409 ± 0.00083 $M_\odot$, $R_B$ = 0.2398 ± 0.0018 $R_\odot$ (Morales et al. 2009; Torres et al. 2010).

Morales et al. (2010) have considered how the presence of cool, dark spots influences the determination of radii from light-curve analysis of eclipsing systems. To do this, they apply a generalized eclipse modelling code to synthetic light curves which include the effects of star-spots. For the case of a distribution of polar spots, they find that their eclipse modelling systematically *overestimates* the sum of the radii of the components by 2–6 per cent. Applying a 3 per cent systematic *decrease* to the radii measured by Morales et al. (2009) for CM Dra, Morales et al. (2010) find that models with an effective dark spot coverage of 17 per cent and a mixing length ratio $\alpha$ = 1 match the observed radii. Based on a fit to the results of Morales et al. (2010), MM12 adopted a relation between radius reduction and dark spot coverage

$$\Delta \log R = -0.0778 f_s. \tag{1.4}$$

Recent determinations of the heavy element abundance for CM Dra are [Fe/H] = −0.30 ± 0.12 (Terrien et al. 2012) and [M/H] = −0.5 ± 0.25 (Kuznetsov et al. 2012). In their fitting of model radii to observed, MM12 assumed solar composition. In light of the new abundance determinations, we have re-analyzed our models of CM Dra A and show in Figure 1 the part of $\delta - f$ space for which fits to the observational data can be obtained for heavy element abundance [Fe/H] = -0.3. (This figure can be compared with Figure 15 of MM12 to see results for solar composition. Note that in the updated models, we still impose an upper limit on the magnetic field strength, $B_{ceiling}$ = 1 MG). The smallest value of $f$ for which a fit is now found is 0.165 and the corresponding value of the magnetic field parameter is $\delta$ = 0.025. The corresponding surface vertical magnetic field strength is 635 G. For [Fe/H] = 0 and $f$ = 0.17, fits were found for $0.005 < \delta < 0.018$.

How much difference to the MM12 results has been caused by using the revised [Fe/H] = −0.3 abundance for CM Dra (instead of the earlier value [Fe/H] = 0)? The most obvious change has been an increase in the surface field strength from 460 − 510 G to 635 G. That is, a decrease in metal abundance by a factor of 2 causes our determination of surface field strength to increase by 25 − 40%. There are no direct measurements of surface magnetic field strengths on CM Dra that we can use for comparison. However, observational data for magnetic cool dwarf stars with masses of 0.2 − 0.4 $M_\odot$ (i.e. a range which includes both components of CM Dra) have been reported by Morin et al (2010): they find surface fields of 400 − 700 G in such stars. This indicates that our magneto-convection models of CM Dra yield surface fields which are consistent with observations, i.e. our approach still yields an acceptable model for the completely convective stars in the CM Dra system.

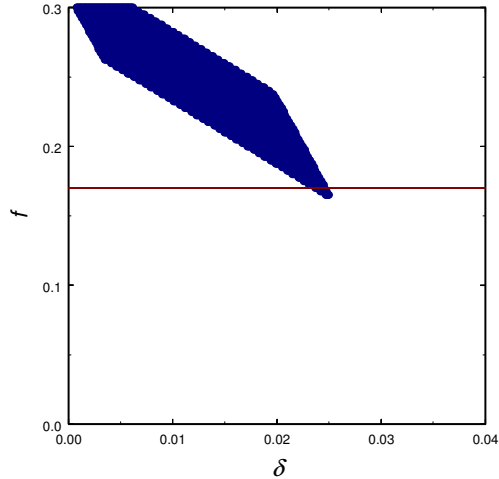

Figure 1. Magnetic models of CM Dra A for [Fe/H] = -0.3. The abscissa refers to the value of $\delta$ in the near-surface layers. Magnetic fields inside the star are nowhere stronger than 1 MG. The figure shows the region of the $\delta$–$f$ plane for which CM Dra A tracks pass through the observed error box. The GT criterion has been used for the onset of convection in the presence of magnetic field. The systematic shift in the observed radii due to spots is assumed proportional to the effective fractional spot coverage, $f$. The horizontal line is placed at $f$ = 0.17 (Morales et al. 2010).

## 4. YY GEMINORUM

YY Geminorum (YY Gem), a member of the Castor Sextuplet, is a double line eclipsing binary, with $P_{orb}$ = 0.814 d, containing two virtually identical M dwarfs. By analyzing the light curve, Torres & Ribas (2002, hereafter TR02) obtained for the mean mass and radius, the values $M$ = 0.5992 ± 0.0047 $M_\odot$, $R$ = 0.6191 ± 0.0057 $R_\odot$. From the stellar colors they determined an effective temperature $T_{eff}$ = 3820 ± 100 K. We find a similar temperature range, $T_{eff}$ = 3750 ± 100 K, by comparing the 2MASS J – H and H – K colors to the predictions of the NextGen atmosphere models. Boyajian et al. (2013) have fitted their derived $T_{eff}$ values for more than 100 stars of spectral types AFGKM by a sixth order polynomial in B-V color. The fit is accurate to a standard deviation of 3.1%. For YY Gem, for which B-V = 1.44 (Gershberg 2002, p658), the polynomial of Boyajian et al (2013) gives $T_{eff}$ = 3786 ± 117 K. By combining these estimates, we adopt $T_{eff}$ = 3775 ± 110 K. The corresponding luminosity is 0.0701 ± 0.0095 $L_\odot$.

TR02 determined the age of the Castor Sextuplet to be 370 ± 40 Myr by applying theoretical isochrones to the two A stars, Castor Aa and Castor Ba. This is the age we adopt for YY Gem. The composition from the theoretical isochrones is found to be close to solar. TR02 point out that at this age the mean component of YY Gem has a radius that is 5 – 15% greater than predicted by standard stellar evolution models. This is an oversizing of at least 5$\sigma$. Another discrepancy from standard evolution models is that YY Gem has a measurable Li abundance (Barrado y Navascués et al. 1997) of $A$(Li) = 12+Log[$N$(Li)/$N$(H)] = 0.11 ± 0.43 whereas standard models predict that Li is destroyed before an age of about 20 Myr. Bonsack (1961) finds Castor Aa has a solar Be/Fe ratio, and so we assume it has a solar system Li abundance, $A$(Li) = 3.28 ± 0.05 (Lodders 2010). Adopting this as the initial Li abundance for YY Gem, its Li depletion is $\Delta A$(Li) = -3.17 ± 0.43.

*4.1. Non-magnetic models applied to YY Gem*

Figure 2 shows the location of the mean component of YY Gem in the log $L$ – Log $R$ diagram. Also shown are a number of standard (non-magnetic) evolutionary tracks ending at age 410 Myr: the "hooks" at the right-hand ends of each track indicate that the star is just settling onto the main sequence after a time of order 400 Myr. The tracks in Fig.2 are for mixing length ratio, $\alpha$, 0.5, 1.0, 1.5 and 2.0. (The curve for $\alpha$ = 0.5 has the lowest luminosity and largest radius.) We see that only the track with $\alpha$ = 0.5 has a "hook" (i.e. at age 410 Myr) which overlaps the observed error box. In Figure 3, we introduce the extra constraint which is provided by the surface Li abundance at age 370 Myr as a function of $\alpha$. The Li measurement is consistent with $\alpha$ values lying between 0.51 and 0.56. Recall that, in order to obtain model fits to the current Sun, the value of $\alpha$ is typically in the vicinity of 1.7 (e.g. $\alpha$ = 1.76 ± 0.08 [Trampedach & Stein 2011], $\alpha$ = 1.684 [MM12]). The much smaller values of $\alpha$ which are required in order to achieve fits to YY Gem (including Li) indicate that the models which fit the YY Gem observations best (at the correct age) are models where the efficiency of convection is clearly smaller than in the Sun.

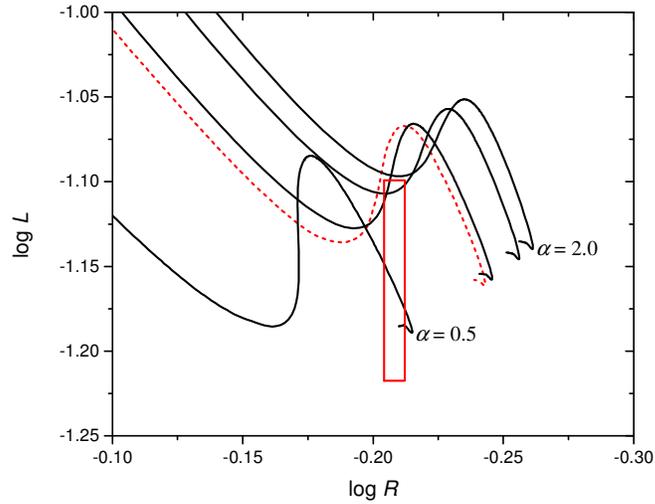

Figure 2. Location of the mean component of YY Gem in the log $L$ – Log $R$ diagram. Also shown are a number of evolutionary tracks ending at age 410 Myr. These tracks are for mixing length ratio, $\alpha$, 0.5, 1.0, 1.5 and 2.0 in order of increasing luminosity and decreasing radii. The dashed line is the evolutionary track for the atmospheric boundary condition.

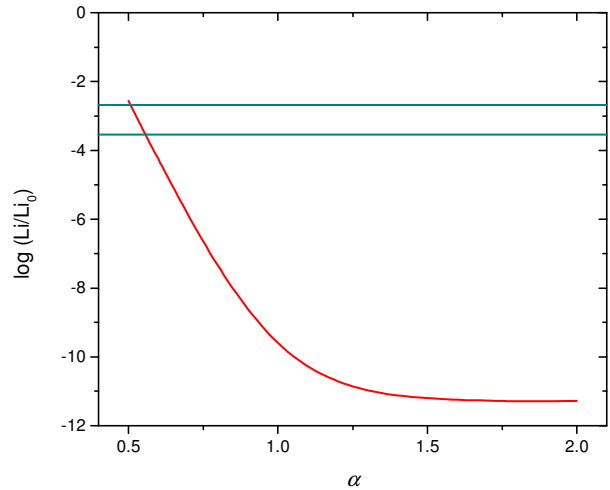

Figure 3. Surface Li depletion in YY Gem models at age 370 Myr as a function of $\alpha$. The observational limits on the Li abundance are denoted by the horizontal lines. Only values of $\alpha$ which lie in the range 0.51 - 0.56 replicate the observed Li abundance.

### 4.2. Models of YY Gem including spots

TR02 modelled the sinusoidal variations of the light curve outside eclipses by means of dark starspots. The resulting dark spot fraction of the mean component is found to be $f_s = 0.03$. Because polar spots and randomly distributed small spots do not affect the light curve, this is strictly a lower limit. In figure 4, we show tracks for different spot fractions, assuming $\alpha = 1$. For this mixing length ratio, spots must occupy more than 40% of the surface area of the star in order to match the observations. By comparing figure 4 with figure 2, it can be seen that there is degeneracy between the effects of varying the spot coverage on stellar properties and the effects of varying the mixing length ratio. For higher mixing length ratios, higher spot fractions are needed. E.g., for the solar calibrated mixing length ratio of $\alpha = 1.684$, a spot fraction of at least 0.7 is needed. Thus, if spots are responsible for the observed properties of YY Gem, the spots must cover more than 40% of the surface area, unless the mixing length ratio is decreased significantly below the solar-calibrated value. The derived range of the Li abundance constrains the spot fraction to be $0.35 < f_s < 0.40$ for $\alpha = 1.0$ and $0.40 < f_s < 0.49$ for $\alpha = 1.7$. From the combined Li abundance and radius constraints, we conclude that the inclusion of spots allows fits with mixing length ratio no larger than 1.0, and for $\alpha = 1.0$, 40% of the surface must be covered by completely dark spots.

Note that, in the case of seeking models of YY Gem, we have not adjusted the radius for polar spots. Unlike the case of CM Dra (Morales et al 2010), inclusion of such spots is not necessary to get good fits to the observations.

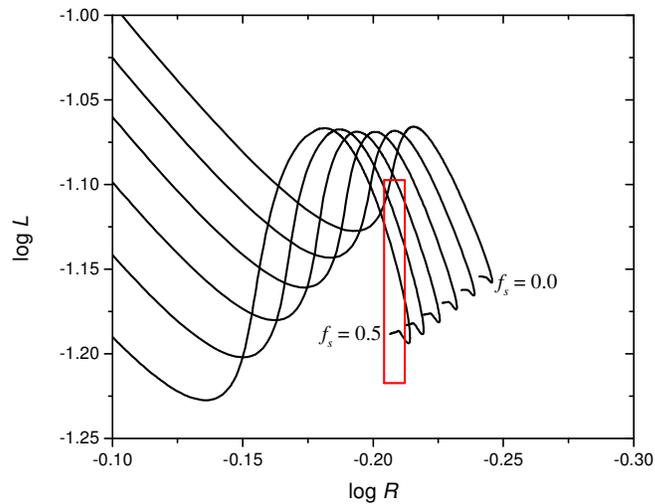

Figure 4. Location of the mean component of YY Gem in the log $L$ – Log $R$ diagram. Also shown are a number of evolutionary tracks for $\alpha = 1.0$ ending at age 410 Myr, for different spot coverages. Spot fractions are, from top to bottom, $f_s = 0.0, 0.1, 0.2, 0.3, 0.4$ and $0.5$.

*4.3. Models of YY Gem based on magneto-convection*

We now consider models in which the effects of inhibition of convection due to the presence of a magnetic field are incorporated by means of the inhibition parameter $\delta$ (see eq. 1.1). In figure 5, tracks are shown for the case in which the magnetic induction has a ceiling of 1 MG. Two sets of tracks are shown, color-coded for $\alpha = 1.0$ (black) and $\alpha = 1.7$ (red).

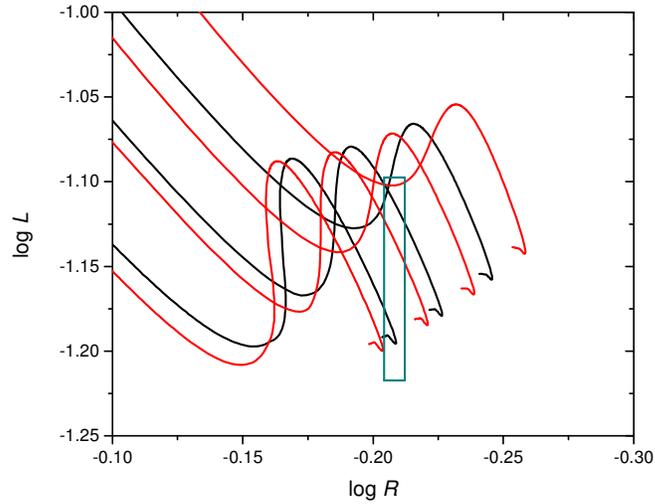

Figure 5. Evolutionary tracks for YY Gem for the magneto-convection model. Results are shown for the case in which the magnetic induction has a ceiling of 1 MG. Tracks are shown for $\alpha = 1.0$ (black) and $\alpha = 1.7$ (red). The magnetic inhibition parameter $\delta$ takes on values of 0.0, 0.01, and 0.02 (black) and values of 0.0, 0.01, 0.02, and 0.03 (red). Increasing $\delta$ values correspond to larger radii and lower luminosities.

Inspection of Fig. 5 indicates that the "hook" on the track with $\delta = 0.02$, $\alpha = 1.0$ is consistent with the observed data. However, we can also get a fit with $\alpha = 1.7$, as long as $\delta$ lies between 0.02 and 0.03. Figure 6 is the same as for figure 5 except that the magnetic field ceiling has been increased to a value of 10 MG. A smaller value of $\delta$ (0.015 for $\alpha = 1.0$; and 0.02 for $\alpha = 1.7$) suffices to cause the "hook" to match the observations in this case.

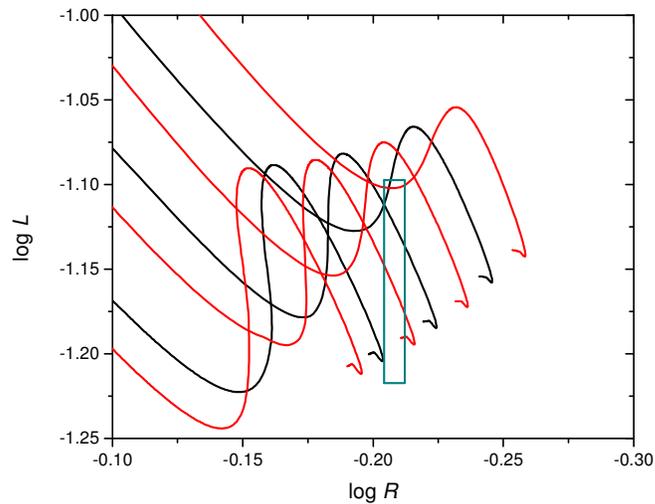

Figure 6. Same as for figure 5 except that the magnetic field ceiling is set to a value of 10 MG. Black curves: $\alpha = 1.0$; red curves: $\alpha = 1.7$.

In Figure 7, we show the results for an even more extreme model in which the ceiling is set equal to 100 MG. (We recognize that it may be difficult for a dynamo mechanism to generate such a field.) Note that there is no longer degeneracy between $\alpha$ and $\delta$: the curves with $\alpha = 1.7$ (red) now lie in positions where *no* interpolation between neighboring red "hooks" falls into the observational error box. In the cases where $\alpha = 1.0$ (black), interpolation between neighboring black "hooks" might just barely fall inside the error box. But to obtain a more confident fit to the box, it appears that $\alpha < 1.0$ would be required. For larger values of $\alpha$, the magneto-convection model therefore sets a limit on the "ceiling" magnetic field: it must be less than 100 MG. As noted above, it seems unlikely that a dynamo could generate fields as strong as this anyway.

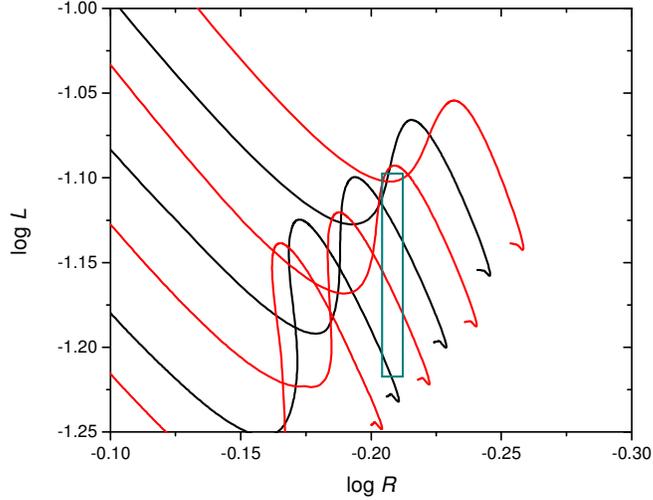

Figure 7. Same as for figure 5 except that the magnetic field ceiling is set equal to 100 MG.

In figure 8 we show a diagram of $B_{ceiling}$ versus $\delta$, in which we illustrate those regions of the diagram where we obtain consistency between our magneto-convection models and three observational quantities: Li abundance, radius, luminosity. All results refer to models which have been evolved to an age of 370 Myr. It is noteworthy that, in the context of our magneto-convection models, regions do exist in the $B_{ceiling} - \delta$ plane where all 3 constraints can be satisfied simultaneously.

For $\alpha = 1.0$ (left-hand panel in Fig. 8), the combination of observed constraints on Li abundance, radius, and luminosity places a firm upper limit on $B_{ceiling}$ of 7.8 MG. This upper limit on $B_{ceiling}$ corresponds to a lower limit of 0.0135 on $\delta$. This value of $\delta$ can be converted (using eq. 1.1 above) to a surface vertical magnetic field: $B_{surf} = \sqrt{4\pi\gamma\delta P_{surf}}$, where $P_{surf}$ is the value of the gas pressure at the photosphere. Inserting the value of $P_{surf}$ from the magneto-convection model at age 370 Myr, this leads to $B_{surf}$ = 250 G. If we limit our models to a canonical value of $B_{ceiling}$ =1 MG, we need $\delta$ values between 0.015 and 0.017, which correspond $B_{surf}$ = 260 – 280 G. At the lowest value of the "ceiling" field which we considered, $B_{ceiling}$ =10 KG, we need $\delta$ values between 0.034 and 0.041 which correspond to $B_{surf}$ = 400 – 440 G.

For $\alpha = 1.7$, the upper limit on $B_{ceiling}$ is 19 MG. The corresponding value of $\delta$ is 0.0195 and the surface field is 310 G. For $B_{ceiling}$ =1 MG, we need $\delta$ values between 0.022 and 0.025, which corresponds to surface fields of 325 – 350 G. For $B_{ceiling}$ =10 KG, we need $\delta > 0.05$, which corresponds to surface fields > 490 G.

Thus, our magneto-convection models predict that the surface magnetic fields in YY Gem are in the range from 250 G to > 490 G, depending on the values assumed for $\alpha$ and $B_{ceiling}$. The most important conclusion from Fig. 8 is that our magneto-convection model is successful in fitting simultaneously three distinct observational constraints on YY Gem. This is a more stringent test of the magneto-convection model than we have previously been able to perform.

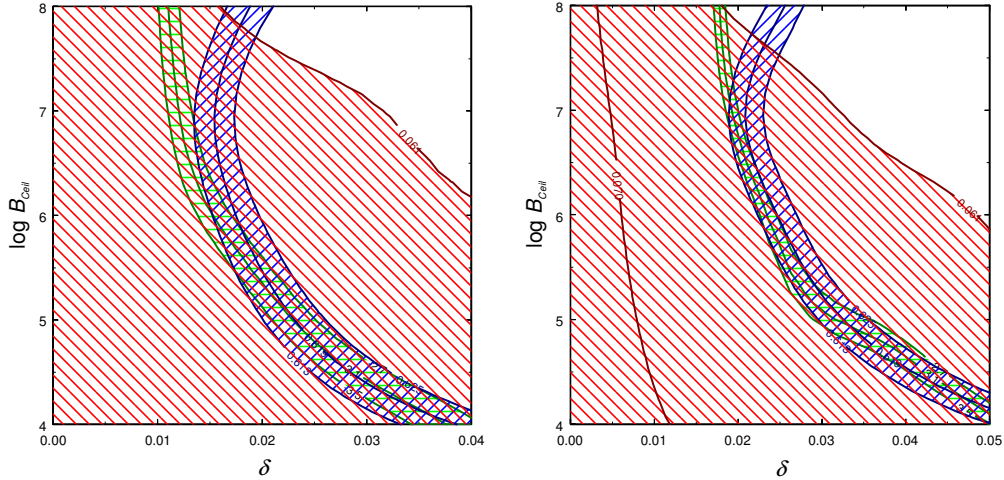

Figure 8. Filled contours of the Li depletion (green), radius (blue) and luminosity (red) at age 370 Myr for $\alpha = 1.0$ (left) and $\alpha = 1.7$ (right). Acceptable simultaneous fits to all 3 observational constraints correspond to where all three shaded regions overlap.

5. CU CANCRI

CU Cancri (CU Cnc) is a M3.5 Ve + M3.5 Ve binary system with $P_{orb}$ =2.77 d. Ribas (2003, hereafter R03) obtained high-precision light curves in multiple photometric bands and by combining his light curve data with radial velocity data from Delfosse et al. (1998) determined the system parameters shown in Table 1. From its space velocity, R03 argues that CU Cnc is a member of the Castor moving group and so, as a starting point for our analysis, we will assume it has the same age and composition as YY Gem. R03 also noted that most evolutionary models underestimate the stellar radii by as much as 10%.

In his high resolution spectra, Ribas reports equivalent widths of ~50 mÅ for the Li I line at $\lambda6708$Å, and based on these, estimates the Li abundance to be log $N$(Li) ≈ −1.1 (in the scale in which log $N$(H) = 12), which we estimate corresponds to a Li depletion $\log[N(Li)/N_0(Li)] = -4.4$. As pointed out by R03, this is in contradiction with Li destruction sequences in clusters and associations (Barrado y Navascués et al. 1999; Stauffer et al. 1999), which indicate that mid-M type stars fully deplete their initial Li abundance in as little as a few times $10^7$ years. In modelling the depletion of Li, we consider in detail only the more massive component CU Cnc A because it has a shallower surface convection zone than CU Cnc B. The presence of Li in CU Cnc B would provide a more challenging constraint on stellar models than its presence in CU Cnc A alone (see section 5.3).

Table 1. Physical properties of CU Cnc

| Component | A | B |
|---|---|---|
| Mass ($M_\odot$) | 0.4333 ± 0.0017 | 0.3980 ± 0.0014 |
| Radius ($R_\odot$) | 0.4317 ± 0.0052 | 0.3908 ± 0.0094 |
| log $g$ (cm s$^{-2}$) | 4.804 ± 0.011 | 4.854 ± 0.021 |
| $T_{eff}$ (K) | 3160 ± 150 | 3125 ± 150 |
| log $L$ (L$_\odot$) | −1.778 ± 0.083 | −1.884 ± 0.086 |
| [Fe/H] | 0.0 | 0.0 |

From the eclipse modelling, R03 finds evidence for dark spots on both components, with the primary having a relatively small spot of radius of 9°, and ~450 K cooler than the photosphere, and the secondary having a larger spot or spot complex of radius of 31° and a temperature difference with the surrounding photosphere of ~200 K. In terms of spot fraction, these parameters correspond to $f_s$(A) = 0.003 and $f_s$(B)= 0.02.

R03 also notes that the absolute magnitude of CU Cnc is dimmer than other stars of the similar mass, and suggests that the apparent faintness of CU Cnc can be explained if its components are some 10% cooler than similar-mass stars or if there is some source of circumstellar dust absorption, possibly from a dusty disk around this relatively young M-type binary.

*5.1. Non-magnetic models of CU Cnc*

Figure 9 shows the location of the two components of CU Cnc in the log $L$ – Log $R$ diagram. Also shown are standard (non-magnetic) evolutionary tracks ending at age 410 Myr, for mixing length parameter $\alpha$ = 0.2, 0.3, 0.5 and 1.0.

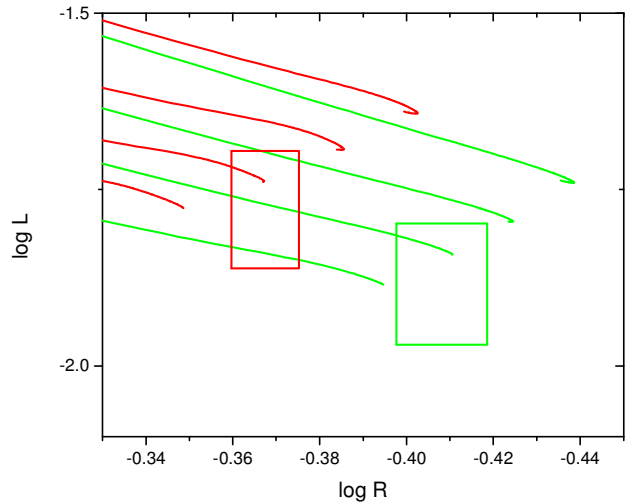

Figure 9. Observational error boxes and evolutionary tracks for CU Cnc A (red) and CU Cnc B (green). Tracks are shown for $\alpha$ = 0.2 (lowest), 0.3, 0.5 and 1.0 (uppermost).

Results in Figure 9 demonstrate how small the values of $\alpha$ must be in order to fit the observed error boxes with standard (non-magnetic) models. For CU Cnc A and B, $0.26 < \alpha < 0.32$, $0.22 < \alpha < 0.38$ respectively. Such values of $\alpha$ are much smaller than the typical values ($\alpha \approx 1.7$) which are required to have models of solar-like stars fit the observed properties. However, they are not the smallest values which have been suggested in the literature for M dwarfs: in the case of the stars Kruger 60A and 60B, with masses of 0.27 and 0.16 $M_\odot$, $\alpha$ values as low as 0.22-0.29 and 0.07-0.17 were obtained by Gabriel (1969) and by Cox et al. (1981) respectively. Such low values indicate a greatly reduced efficiency of convective transport relative to the "typical" conditions in a sun-like star. In this regard, it is noteworthy that Kruger 60B is a flare star (Gershberg 2002, p. 649), and therefore has active magnetic fields which are being stressed by the convective motions.

For values of $\alpha$ consistent with the radius and luminosity constraints, the surface Li abundance is reduced to below detectable limits in less than 50 Myr. The presence of Li would rule out models for CU Cnc in which the radius inflation is due solely to reduced mixing length ratio.

*5.2. Models of CU Cnc with spots*

Figure 10 is similar to Fig. 4 except for CU Cnc. Fig. 10a (left) refers to $\alpha = 1.0$ and Fig. 10b (right) is for $\alpha = 1.7$. The $f_s$ values are 0.0 (topmost), 0.1, 0.2, 0.3, 0.4 and 0.5 (lowest). To match the data in Fig. 10a requires $0.39 < f_s < 0.49$ and $0.28 < f_s < 0.50$ for components A and B, respectively. In both cases, the spotted models have interior structure which causes Li to be depleted to below detectable levels in stars with ages of 410 Myr. To match the data in Fig. 10b requires $0.40 < f_s < 0.52$ and $0.33 < f_s < 0.53$ for components A and B, respectively. Again, in the presence of larger $\alpha$, spotted models of CU Cnc have Li depleted to below detectable levels.

These attempts to apply the Li constraint to spotted models of CU Cnc A and B point to the following conclusion: the detection of Li rules out models where the only effects of magnetic fields are to cause dark spots to occur on the surface. We also note that the spot coverages required to match the radius and luminosity data are significantly higher than those from the spot modelling of Ribas (2013).

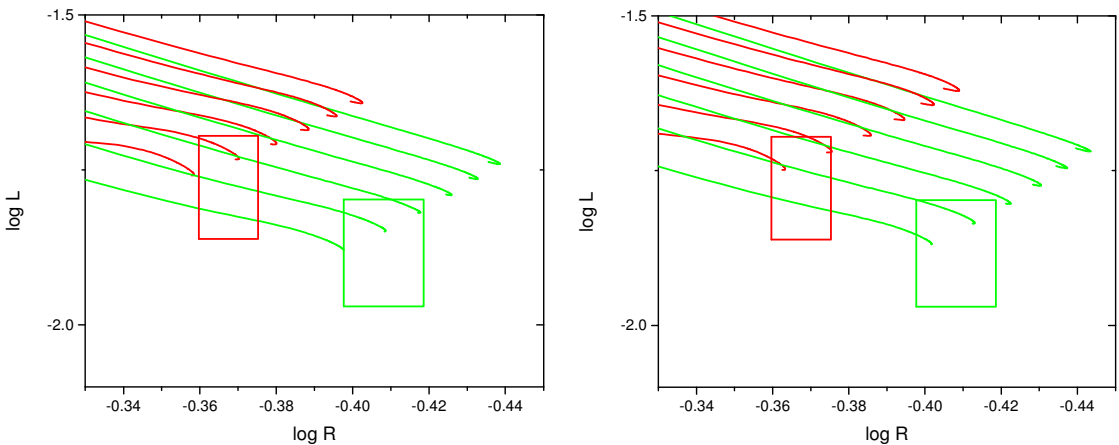

Figure 10. Locations of CU Cnc A (red box) and CU Cnc B (green box) in the radius – luminosity diagram.. Also shown are evolutionary tracks for CU Cnc A (red) and B (green) when the surfaces contain dark spots, assuming $\alpha = 1.0$ (left) and $\alpha = 1.7$ (right). Spot fractions are, from top to bottom, $f_s = 0.0, 0.1, 0.2, 0.3, 0.4$ and $0.5$.

## 5.3. Models of CU Cnc with magneto-convection

In figures 11, 12 and 13 are shown results for magneto-convection models with ceilings of 1, 10, 100 MG respectively. The symbols indicate where the Li abundance is A(Li) = -0.6 (stars) and -1.6 (circles).

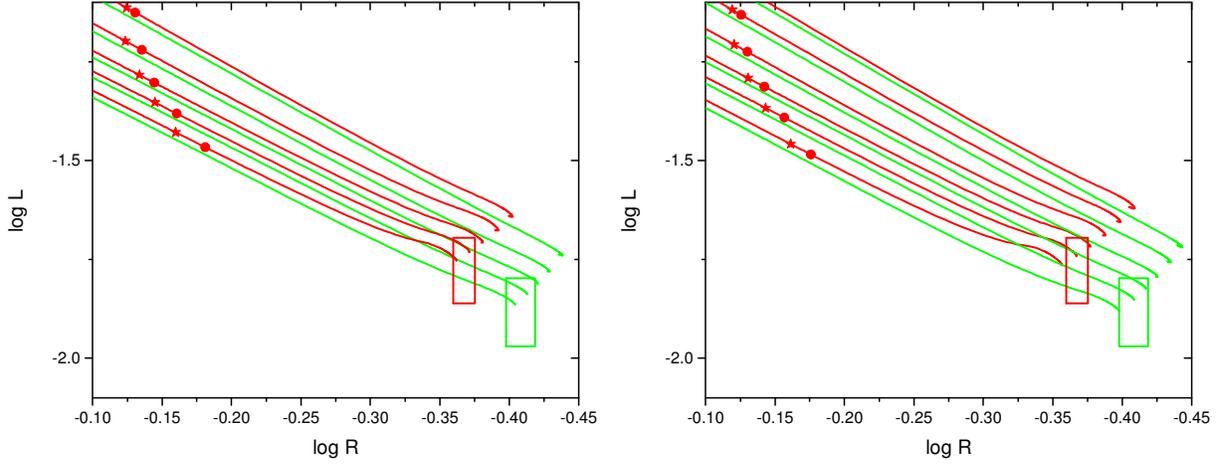

Figure 11. Magneto-convection models for CU Cnc A (red) and CU Cnc B (green) compared to observational data (boxes). $B_{ceiling}$ = 1 MG in this figure. Left panel: $\alpha$ = 1.0. Right panel: $\alpha$ = 1.7. For each track, the magnetic inhibition parameter $\delta$ takes on a different value: 0 (topmost track), 0.01, 0.02, 0.03, 0.04 and, in right panel only, 0.05. Each evolution track is terminated at an age of 410 Myr. If the age of CU Cnc is equal to that of the Castor system, then the acceptable tracks as regards $L$ and $R$ are those where the endpoints for both components lie simultaneously inside the corresponding box. In order to satisfy also the Li constraint, the boxes should also lie between a filled star and a filled circle.

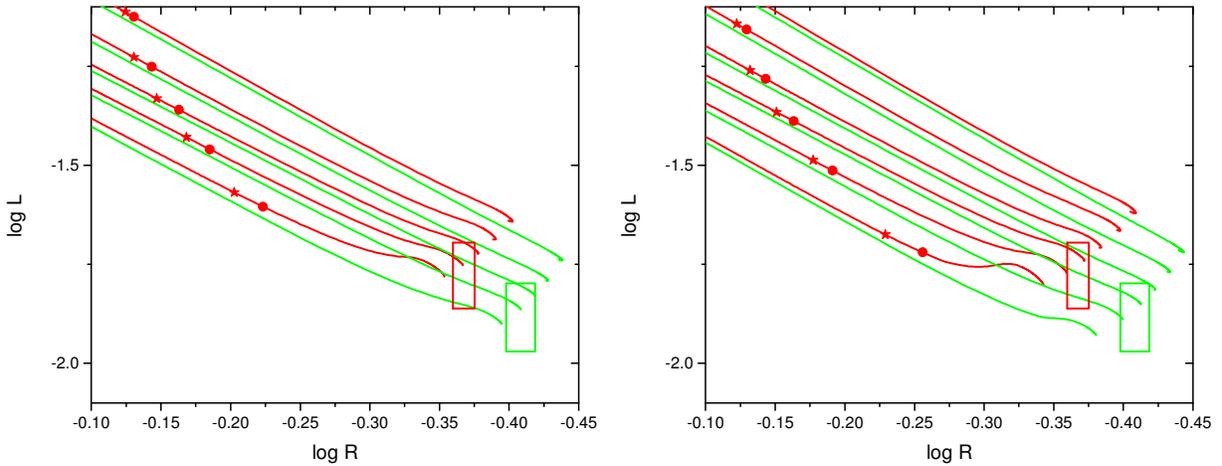

Figure 12. Magneto-convection models for CU Cnc A (red) and CU Cnc B (green) compared to observational data (boxes). $B_{ceiling}$ = 10 MG in this figure. Left panel: $\alpha$ = 1.0. Right panel: $\alpha$ = 1.7. For each track, the magnetic

inhibition parameter $\delta$ takes on a different value: 0 (topmost track), 0.01, 0.02, 0.03, 0.04, and in right panel only 0.05. Each evolution track is terminated at an age of 410 Myr. If the age of CU Cnc is equal to that of the Castor system, then the acceptable tracks as regards $L$ and $R$ are those where the endpoints for both components lie simultaneously inside the corresponding box. In order to satisfy also the Li constraint, the boxes should also lie between a filled star and a filled circle.

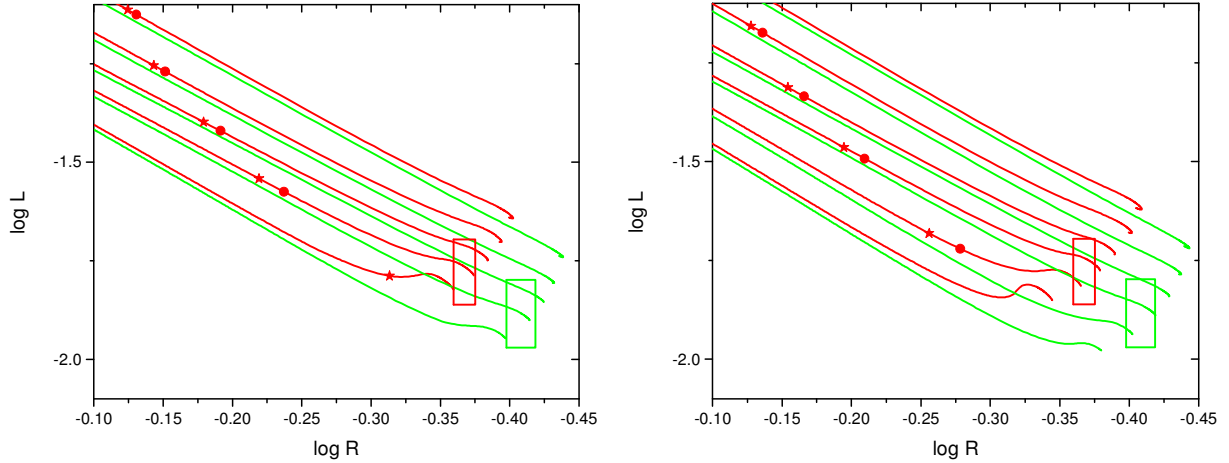

Figure 13. Magneto-convection models for CU Cnc A (red) and CU Cnc B (green) compared to observational data (boxes). $B_{ceiling} = 100$ MG in this figure. Left panel: $\alpha = 1.0$. Right panel: $\alpha = 1.7$. For each track, the magnetic inhibition parameter $\delta$ takes on a different value: 0 (topmost track), 0.01, 0.02, 0.03, 0.04 and in right panel only 0.05. Each evolution track is terminated at an age of 410 Myr. If the age of CU Cnc is equal to that of the Castor system, then the acceptable tracks as regards $L$ and $R$ are those where the endpoints for both components lie simultaneously inside the corresponding box. In order to satisfy also the Li constraint, the boxes should also lie between a filled star and a filled circle.

Inspection of Figs. 11-13 indicates that for $\alpha = 1.0$, the only tracks which are consistent with all three constraints (Li, radius, luminosity) at an age of 410 Myr are those with $\delta = 0.04$ and $B_{ceiling} = 100$ MG.

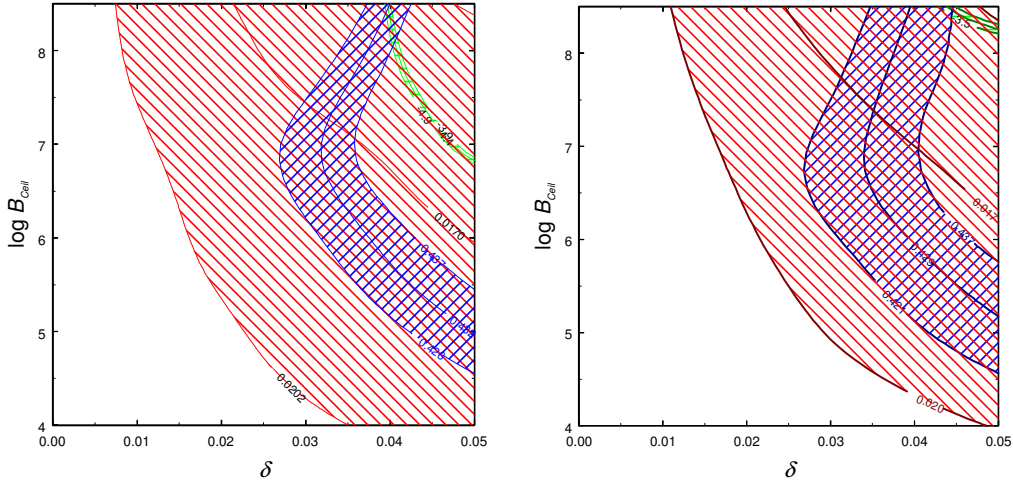

Figure 14. Plot of the $B_{ceiling} - \delta$ plane for magneto-convection models of CU Cnc A at an age of 400 Myr, as constrained by observed values of Li depletion (green), radius (blue) and luminosity (red).

In Figure 14, we present results analogous to those in Fig. 8 except for CU Cnc A. For $\alpha = 1.0$, the Li abundance, $R$ and $L$ measurements require $B_{ceiling} > 100$ MG. For $\alpha = 1.7$, $B_{ceiling} > 300$ MG is needed. These ceiling fields are so large that it is difficult to see how dynamo operation can produce them. In the event that a dynamo cannot produce such large fields, we are forced to conclude that our magneto-convection model *cannot* simultaneously satisfy all three constraints of Li abundance, $R$, and $L$ for CU Cnc A.

## 6. SUMMARY AND DISCUSSION

In this paper, we report on results obtained by applying our magneto-convection model to CM Dra, YY Gem, and CU Cnc, three M dwarf eclipsing binaries for which the most precise radii, masses, and luminosities are available. Our magneto-convection model (introduced in MM01), is based on the criterion for thermal convection in the presence of a uniform vertical magnetic field derived by Gough & Tayler (1966).

In light of new determinations of its heavy element abundance, we have revisited the explanation of the inflated radii of the components of the M-dwarf binary CM Dra as due to magnetic inhibition of convection. Our results for CM Dra show that, once the observed radii are corrected due to the effect of the (assumed) presence of dark spots on eclipse modelling, surface vertical magnetic field components of strength 635 G are sufficient to inflate the model radius of CM Dra A to match its observed radius. Surface fields with such strengths are consistent with observational data obtained by Morin et al (2010) for stars with masses equal to those in CM Dra.

We have also applied our model to the M dwarf binaries YY Gem and CU Cnc. The detection of Li in these systems allows us to place new constraints on our model, especially as regards the value of an assumed "ceiling" magnetic field strength.

For the YY Gem mean component, we find that the ceiling cannot exceed about 8 MG if the mixing length ratio $\alpha = 1.0$, or 20 MG for our solar calibrated value $\alpha = 1.7$. Are fields as strong as this plausible in the deep interior of YY Gem? To address this, we ask: how large might the magnetic fields be which are generated by dynamo operation at the base of the YY Gem convection zone? Assuming that both YY Gem components are tidally locked with the orbital period ($P$ = 0.814 days), the angular velocity is 8.93 $10^{-5}$ rad s$^{-1}$, and the rotational velocity $v_r$ at the base of the convection zone is 2.6 $10^6$ cm s$^{-1}$. The density $\rho$ at the base of the convection zone is 1.5 g cm$^{-3}$. If we assume equipartition between the kinetic energy density of rotation $0.5\rho v_r^2$ and the magnetic energy density $B^2/8\pi$, we find $B$ = 11 MG in YY Gem. In general models of distributed dynamos (Browning 2008), the maximum $B$ values are no more than 2 or 3 times the equipartition value. Therefore, requiring field strengths to be as large as the "ceiling" value of 8 MG in YY Gem (as we have found in our magneto-convection models with $\alpha$ = 1.0: see Fig. 8a), does not appear implausible in the context of dynamo operation in the YY Gem system.

In contrast, we find for CU Cnc A that the ceiling field must exceed 100 MG, otherwise Li would be depleted below the observed value. Is a field strength of 100 MG at the base of the surface convection reasonable? Assuming that CU Cnc A is tidally locked with its orbit, its rotation rate is 2.63 $10^{-5}$ rad s$^{-1}$, and $v_r$ at the base of the convection zone is 5.4 $10^5$ cm s$^{-1}$. The density there is 5.3 g cm$^{-3}$. Again assuming $0.5\rho v_r^2 = B^2/8\pi$, we find that $B$ in CU Cnc A is 4 MG. Even if we amplify this by factors of 2-3 (Browning 2008), it seems highly unlikely that fields of 100 MG could be dynamo generated in CU Cnc A. We note that in the context of a different model, Feiden & Chaboyer (2013) have also found that very strong fields are required to account for observational constraints in CU Cnc.

We suggest a number of possible resolutions to the CU Cnc problem:

(1) Perhaps the reported Li abundance determination is incorrect. If in fact Li is completely depleted, we find that surface fields as low as 420 G are sufficient to give the required radius inflation. Such fields would fit perfectly into the range of observed surface fields reported by Morin et al. (2010) for stars in the mass range 0.2 - 0.4 $M_\odot$ (such as CU Cnc B: and CU Cnc A almost falls into this range).

(2) CU Cnc might not actually be a member of the Castor moving group or the Castor moving group is not comprised of objects formed in the same place and at the same time (Mamajek et al. 2013). If this is the case, then the age and composition of CU Cnc are not well constrained. For our non-magnetic solar composition models of CU Cnc A, the observed degree of Li depletion for a star with a mass of 0.43 $M_\odot$ is obtained at ages of no more than 13 – 15 Myr. At such a young age, the radius is about 0.75 $R_\odot$, which is significantly *larger* than the observed radius. Reducing the heavy element abundance does not make a major difference. For example, if [Fe/H] = -1, the radius is 0.61 $R_\odot$ at the lithium age. It seems inescapable that the age of CU Cnc A cannot be as young as 13-15 Myr. The observed radii and luminosities suggest that the components are on, or very close to, the main sequence. In such a case, very large interior fields are needed to suppress Li depletion.

(3) Perhaps the Li on the surface is in the process of being accreted from a circumstellar disk. In the absence of Li depletion, this would require a mean accretion rate of 2 $10^{-14}$ $M_\odot$ yr$^{-1}$. Simulations which include accretion of solar system composition material onto our $\delta$ = 0.04 model, $B_{ceiling}$ = 1 MG model at a uniform rate show that the accretion rate must be between 2 $10^{-13}$ $M_\odot$ yr$^{-1}$ and 2 $10^{-12}$ $M_\odot$ yr$^{-1}$ to balance the destruction of Li by proton capture and maintain the Li abundance at the observed level. If accretion were from a gaseous circumstellar disk, the disk has to have had an initial mass greater than about 7 $10^{-5}$ $M_\odot$ = 2 $10^{-4}$ $M_*$. Alternatively the Li could be supplied by accretion of a single object. For an object that is mainly rocky material, its mass needs to be approximately that of the planet Mercury.

None of the stars in the three systems considered here have direct measurement of their surface magnetic fields, but we can get some idea of the expected field strengths from the observed correlation of 'magnetic flux' (more precisely the integral of the scalar field strength over the stellar surface) with X-ray luminosity (Fisher et al 1998; Pevtsov et al. 2003; Feiden & Chaboyer 2013).

Feiden & Chaboyer (2013) have used data from the ROSAT All-Sky Survey Bright Source Catalogue (Voges et al. 1999) to determine X-ray fluxes of YY Gem and CU Cnc. Feiden & Chaboyer point out these are likely upper limits on the X-ray fluxes because of source confusion due to ROSAT having relatively poor spatial resolution. The data for YY Gem are contaminated by any X-rays emanating from Castor A and B and the data for CU Cnc by X-rays from the active M dwarf binary CV Cnc. Since X-rays from Castor A and B are most likely from the M dwarf companions rather than the A stars, we have divided the observed X-ray flux by 4 to give an estimate of the X-ray flux from the mean YY Gem component. To determine the X-ray luminosity of the YY Gem mean component, we use the parallax determined by TR02 for Castor. We have followed the procedure of Feiden & Chaboyer (2013) to determine the X-ray luminosity per star of CM Dra. The distances to CM Dra and CU Cnc were obtained from the parallaxes given by Harrington & Dahn (1980), and van Leeuwen (2007), respectively. Table 2 compares the magnetic flux predicted by our $\alpha = 1.7$, 1 MG ceiling models with the values from the X-ray luminosity – magnetic flux relation of Feiden & Chaboyer (2013).

Table 2. Comparison of predicted magnetic flux with that from X-ray luminosity – magnetic flux relation

| System | $X_{cr}$ (counts s$^{-1}$) | Hardness ratio | $\pi$ (mas) | $L_X$ ($10^{28}$ erg s$^{-1}$) | Log $\Phi_{FC}$ (Mx) | Log $\Phi_{pred}$[1] (Mx) |
|---|---|---|---|---|---|---|
| CM Dra | 0.18 ± 0.02 | -0.30 ± 0.07 | 68 ± 4 | 1.55 ± 0.52 | 24.81 ± 0.52 | 24.40 ± 0.06 |
| YY Gem | 3.70 ± 0.09 | -0.15 ± 0.02 | 66.90 ± 0.63 | 18.6 ± 1.1 | 25.32 ± 0.48 | 24.93 ± 0.06 |
| CU Cnc | 0.73 ± 0.05 | -0.14 ± 0.06 | 90.37 ± 8.22 | 2.03 ± 0.63 | 24.87 ± 0.51 | 24.75 ± 0.04 |

[1] For $\alpha = 1.7$, $B_{ceiling} = 1$ MG.

We see that our predicted fluxes for CM Dra, YY Gem and CU Cnc are all consistent with the values from the $L_X - \Phi$ relation. In the GT model, only the vertical component of the magnetic field is responsible for magnetic inhibition of convection. Hence our flux estimates are strictly lower limits. If the magnetic field direction is randomly distributed, the total flux would be higher by $\sqrt{3}$, which in each case remains within the error bars of the $L_X - \Phi$ relation. In contrast, Feiden & Chaboyer (2013) find that the magnetic field strengths from their model for magneto-convection (Feiden & Chaboyer 2012) needed to fit the observed radii of the components of the eclipsing systems UV Psc, YY Gem and CU Cnc are higher than those predicted by the $L_X - \Phi$ relation. A direct measurement of the surface magnetic field in at least one of the eclipsing systems would be very useful in discriminating between modelling approaches.

Acknowledgements. D.J.M. thanks DE Space Grant for partial support of this work.